\documentstyle[12pt]{article}
\begin{document}

\author{Amand Faessler$^1$, A. J. Buchmann$^1$, M. I. Krivoruchenko$^{1,2}$ \\
%EndAName
{\small $^1${\it Institut f\"ur Theoretische Physik, Universit\"at
T\"ubingen, Auf der Morgenstelle 14 }}\\
{\small {\it D-72076 T\"ubingen, Germany}}\\
{\small $^2${\it Institute for Theoretical and Experimental Physics,
B.Cheremushkinskaya 25}}\\
{\small {\it 117259 Moscow, Russia}}}
\title{Constraints to Coupling Constants of the $\omega $- and $\sigma $-Mesons
with Dibaryons}
\date{}
\maketitle

\begin{abstract}
The effect of narrow dibaryon resonances to nuclear matter and structure of
neutron stars is investigated in the mean-field theory (MFT) and in the
relativistic Hartree approximation (RHA). The existence of massive neutron
stars imposes constraints to the coupling constants of the $\omega $- and $%
\sigma $-mesons with dibaryons. We conclude that the experimental candidates
to dibaryons d$_1$(1920) and d'(2060) if exist form in nuclear matter a Bose
condensate stable against compression. This proves stability of the ground
state for nuclear matter with a Bose condensate of the light dibaryons.
\end{abstract}

\newpage 

The prospect to observe the long-lived H-particle predicted in 1977 by R.
Jaffe \cite{Jaf} stimulated considerable activity in the experimental
searches of dibaryons. It was proposed to examine the H-particle production
in different reactions \cite{San}. The experiments \cite{Aok} did not give a
positive sign for the H-particle, however, the existence of the H-particle
remains an open question which must eventually be settled by experiment. The
non-strange dibaryons with exotic quantum numbers, which have a small width
due to zero coupling to the $NN$-channel, are promising candidates for
experimental searches \cite{Mul}. The data on pion double charge exchange
(DCE) reactions on nuclei \cite{Bil} exhibit a peculiar energy dependence,
which can be interpreted \cite{Mar} as evidence for the existence of a
narrow d' dibaryon with quantum numbers $T=0$, $J^p=0^{-}$ and the total
resonance energy of $2063$ MeV. Recent experiments at TRIUMPF (Vancouver)
and CELSIUS (Uppsala) seem to support the existence of the d' dibaryon \cite
{Mey}. A method for searching narrow, exotic dibaryon resonances in the
double proton-proton bremsstrahlung reaction is discussed in Ref. \cite{Ger}%
. Recently, some indications for a d$_1$(1920) dibaryon in this reaction
have been found \cite{Khr}.

When density of nuclear matter is increased beyond a critical value,
production of dibaryons becomes energetically favorable. Dibaryons are Bose
particles, so they condense in the ground state and form a Bose condensate 
\cite{Bal,Kri}. An exactly solvable model for a one-dimensional Fermi-system
of fermions interacting through a potential leading to a resonance in the
two-fermion channel is analyzed in Ref. \cite{Buc}. The behavior of the
system with increasing the density can be interpreted in terms of a Bose
condensation of two-fermion resonances. The effect of narrow dibaryon
resonances on nuclear matter in the mean field theory (MFT) is analyzed in
Refs. \cite{Fae,Faes}. In the limit of vanishing decay width, a dibaryon
can be approximately described as an elementary field.

Despite the dibaryon Bose condensate does not exist in ordinary nuclei,
dibaryons affect properties of nuclear matter and the ordinary nuclei
through a Casimir effect. Presence of the background $\sigma $-meson mean
field inside of nuclei modifies the nucleon and dibaryon masses and in turn
modifies the zero-point vacuum fluctuations of the nucleon and dibaryon
fields. This effect contributes to the energy density and pressure. It can
be evaluated within the relativistic Hartree approximation (RHA). For
nucleons, this effect is well known \cite{Wal}. In the loop expansion of 
  quantum hadrodynamics (QHD), 
MFT corresponds to the lowest approximation (no loops), while RHA corresponds 
to the one-loop approximation in a calculation of the equation 
of state for nuclear matter.

At zero temperature, a uniformly distributed system of bosons with
attractive potential is energetically unstable against compression and
collapses \cite{Abr}. In such a case, the long wave excitations (sound in
the medium) have imaginary dispersion law: The square of the sound velocity
is negative $a_s^2<0$. The amplitude of these excitations increases with the
time, providing instability of the system. It is necessary to analyze
dispersion laws of other elementary excitations also. We shall see, however,
that in MFT and RHA only sound waves can generate an instability. The ground
state of nuclear matter with a Bose condensate of dibaryons is stable or
unstable against small perturbations according as the repulsive $\omega $%
-meson exchange or the attractive $\sigma $-meson exchange is dominant
between dibaryons.

In this paper, we investigate the hypothesis that the dibaryon matter is
unstable against compression. In such a case, formation of dibaryons in
nuclear matter can be treated as a possible mechanism for a phase transition
into the quark matter. If central density of a massive neutron stars exceeds
a critical value for formation of dibaryons, the neutron star should convert
into a quark star, a strange star, or a black hole. Some of the observed
pulsars are identified quite reliably with ordinary neutron stars \cite{Sha}%
. From the requirement that the dibaryon formation is not energetically
favorable at densities lower than the central density of neutron stars with
a mass $1.3M_{\odot }$, we derive constraints to the coupling constants of
the mesons and dibaryons d$_1$(1920) and d$^{\prime }$(2060) and conclude
that narrow dibaryons in this mass range can form a Bose condensate stable
against perturbations only. The effect of the dibaryons to stability and
structure of neutron stars in different phenomenological models is analyzed
in Refs. \cite{Kri,Tam}. Constraints to the binding energy of strange matter 
\cite{Wit} from the existence of massive neutron stars are discussed in Ref. 
\cite{KrMa}.

The dibaryonic extension of the Walecka model \cite{Wal} is obtained by
including dibaryons to the Lagrangian density \cite{Fae,Faes} 
\begin{equation}
\label{I}
\begin{array}{c}
{\cal L}=\bar \Psi (i\partial _\mu \gamma _\mu -m_N-g_\sigma \sigma
-g_\omega \omega _\mu \gamma _\mu )\Psi +\frac 12(\partial _\mu \sigma
)^2-\frac 12m_\sigma ^2\sigma ^2 \\ -\frac 14F_{\mu \nu }^2+\frac 12m_\omega
^2\omega _\mu ^2+(\partial _\mu -ih_\omega \omega _\mu )\varphi
^{*}(\partial _\mu +ih_\omega \omega _\mu )\varphi -(m_D+h_\sigma \sigma
)^2\varphi ^{*}\varphi . 
\end{array}
\end{equation}
Here, $\Psi $ is the nucleon field, $\omega _\mu $ and $\sigma $ are fields
of the $\omega $- and $\sigma $-mesons, $F_{\mu \nu }=\partial _\nu \omega
_\mu -\partial _\mu \omega _\nu $, $\varphi $ is the dibaryon
isoscalar-scalar (or isoscalar-pseudoscalar) field. The values $m_\omega \ $
and $m_\sigma $ are the $\omega $- and $\sigma $-meson masses and the values 
$g_\omega $, $g_\sigma $, $h_\omega $, $h_\sigma $ are coupling constants of
the $\omega $- and $\sigma $-mesons with nucleons ($g$) and dibaryons ($h$).

The $\sigma $-meson mean field $\sigma _c$ determines the effective nucleon
and dibaryon masses in the medium 
\begin{equation}
\label{II}m_N^{*}=m_N+g_\sigma \sigma _c, 
\end{equation}
\begin{equation}
\label{III}m_D^{*}=m_D+h_\sigma \sigma _c. 
\end{equation}

The nucleon scalar density in the RHA is defined by expression \cite{Wal}

\begin{equation}
\label{IV}\rho _{NS}=<\bar \Psi (0)\Psi (0)>=\gamma \int \frac{d{\bf p}}{%
(2\pi )^3}\frac{m_N^{*}}{E^{*}({\bf p})}\theta (p_F-|{\bf p}|)-4m_N^3\zeta
(m_N^{*}/m_N) 
\end{equation}
where

$$
4\pi ^2\zeta (x)=x^3lnx+1-x-\frac 52(1-x)^2+\frac{11}2(1-x)^3. 
$$
The last term in Eq.(4) occurs after the renormalization of the scalar
density. Here, $\gamma =2$ for neutron matter and $\gamma =4$ for nuclear
matter.

We investigate here properties of the nuclear matter below the critical
density for formation of dibaryons, so the dibaryon condensate is zero $%
<|\varphi (0)|>=0$. The vacuum contribution to the scalar density of
dibaryons can be found to be 
\begin{equation}
\label{V}2m_D^{*}\rho _{DS}=2m_D^{*}<\varphi (0)^{*}\varphi (0)>=m_D^3\zeta
(m_D^{*}/m_D). 
\end{equation}
It differs from the nucleon term in the sign and in the statistical factor
(one should replace $4$ ($=2_s\times 2_I)\rightarrow 1$). The
self-consistency condition for the nucleon effective mass has the form 
\begin{equation}
\label{VI}m_N^{*}=m_N-\frac{g_\sigma }{m_\sigma ^2}(g_\sigma \rho
_{NS}+h_s2m_D^{*}\rho _{DS}). 
\end{equation}

The renormalized vacuum contribution to the nucleon energy-momentum tensor
has the form \cite{Wal} 
\begin{equation}
\label{VII}<T_{\mu \nu }^N(0)>_{vac}=-4g_{\mu \nu }m_N^4\eta (m_N^{*}/m_N) 
\end{equation}
where%
$$
16\pi ^2\eta (x)=x^4lnx+1-x-\frac 72(1-x)^2+\frac{13}3(1-x)^3-\frac{25}{12}%
(1-x)^4. 
$$

For dibaryons, we get expression 
\begin{equation}
\label{VIII}<T_{\mu \nu }^D(0)>_{vac}=g_{\mu \nu }m_D^4\eta (m_D^{*}/m_D). 
\end{equation}

The elementary excitations in nuclear matter with a Bose condensate of
dibaryons correspond to nucleons and antinucleons, $\omega $-mesons, $\sigma 
$-mesons, and dibaryons and antidibaryons. The dispersion laws for these
quasiparticles are found in Ref. \cite{Faes}. The nucleon and antinucleon
dispersion laws have the same form as in the vacuum with a replacement $%
m_N\rightarrow m_N^{*}$ and therefore cannot generate an instability of the
system. The dispersion laws of $\omega $-mesons, $\sigma $-mesons, and
antidibaryons turn out to be real also. The possible source of the
instability are the dibaryon quasiparticle excitations only, which are
responsible for a long wave perturbations of the system and connected with
existence of the sound in the medium.

The square of the sound velocity has the form \cite{Faes} 
\begin{equation}
\label{IX}a_s^2=\frac \alpha {1+\alpha } 
\end{equation}
where 
\begin{equation}
\label{X}\alpha =2\rho _{DS}\frac{m_\sigma ^2}{\tilde m_\sigma ^2}(\frac{%
h_\omega ^2}{m_\omega ^2}-\frac{h_\sigma ^2}{m_\sigma ^2}) 
\end{equation}
and $\tilde m_\sigma ^2=m_\sigma ^2+2h_\sigma ^2\rho _{DS}$. We see that $%
a_s^2>0$ for 
\begin{equation}
\label{XI}\frac{h_\omega ^2}{m_\omega ^2}>\frac{h_\sigma ^2}{m_\sigma ^2}. 
\end{equation}
The validity of inequality (11) is sufficient condition for stability of the
ground state of nuclear matter with a Bose condensate of dibaryons.

The physical meaning of the inequality (11) can be clarified by considering
the interaction energy of a uniformly distributed dibaryon matter $\rho
_{DV}({\bf x}_1)=\rho _{DV}=$ constant: 
\begin{equation}
\label{XII}W=\frac 12\int d{\bf x}_1d{\bf x}_2\rho _{DV}({\bf x}_1)\rho
_{DV}({\bf x}_2)V(|{\bf x}_1-{\bf x}_2|). 
\end{equation}
The Yukawa potential $V(r)$ for two dibaryons has the form 
\begin{equation}
\label{XIII}V(r)=\frac{h_\omega ^2}{4\pi }\frac{e^{-m_\omega r}}r-\frac{%
h_\sigma ^2}{4\pi }\frac{e^{-m_\sigma r}}r. 
\end{equation}
The integration gives 
\begin{equation}
\label{XIV}W=\frac 12N_D\rho _{DV}(\frac{h_\omega ^2}{m_{_\omega }^2}-\frac{%
h_\sigma ^2}{m_{_\sigma }^2}) 
\end{equation}
where $N_D$ is the number of dibaryons. When the dibaryon density $\rho
_{DV} $ increases, the energy for $a_s^2>0$ increases also, the pressure is
positive, and so the system is stable.

At present, the coupling constants of the mesons with dibaryons are not
known with precision good enough to draw a definite conclusion concerning
the stability of the dibaryon matter.

Here, we show that violation of the inequality (11) for light dibaryons is
in contradiction with the existence of massive neutron stars.

Given that the ratio $h_\sigma /(2g_\sigma )$ between the $\sigma $-meson
couplings with dibaryons and nucleons is fixed, one can find the $\omega $-
and $\sigma $-meson couplings with nucleons by fitting the nuclear matter
binding energy $E/A-m_N=-15.75$ MeV at the empirical equilibrium density $%
\rho _0=0.148$ fm$^{-3}$ determined from the density in the interior of $%
^{208}Pb$ \cite{Wal}. It corresponds to the equilibrium Fermi wavenumber $%
k_F=1.3$ fm$^{-1}$. The dependence of the effective nucleon mass at the
equilibrium density on the ratio $h_\sigma /(2g_\sigma )$ is shown on Fig.1
(a). In Figs.1 (b) and (c) we give dependence of the incompressibility $%
K=9\rho _0(\partial ^2\varepsilon /\partial \rho ^2)|_{\rho =\rho _0}$ and
the asymmetry coefficient $a_4$ on the ratio $h_\sigma /(2g_\sigma )$. In
Fig.1 (d), the values $C_s^2=g_\sigma ^2(m_N/m_\sigma )^2$ and $C_\omega
^2=g_\omega ^2(m_N/m_\omega )^2$ are plotted. Notice that $h_\sigma
/(2g_\sigma )=0$ is equivalent to RHA with no dibaryons. For comparison, we
give the results of MFT where the effect of dibaryons to the nuclear matter
below the critical density for occurrence of a Bose condensate of dibaryons
is absent. The results RHA for different dibaryons are not much different. 
  In this work, we study the physical implications of the effective
Lagrangian (1) describing nucleon and dibaryon degrees of
freedom. Other baryons can be included in the QHD framework in
a similar way. Their effect is quite small. We have checked that the 
Casimir effect caused by the inclusion
of other octet baryons ($2 \times 8 = 16$ degrees of freedom) shift the 
critical value $h_{\sigma }$ only by about 25$\%$ if one assumes that the
${\sigma }$-meson is an SU(3)$_{f}$ singlet.
The inclusion of octet and
decuplet baryons ($2 \times 8 + 4 \times 10 = 56$ degrees of freedom) with 
a universal sigma-meson coupling constant increases the critical value of
$h_{\sigma }/(2g_{\sigma })$ to 1.2.

When the ratio $h_\sigma /(2g_\sigma )$ approaches a value $0.8$, the system
of equations blows up and the empirical equilibrium properties of the
nuclear matter can no longer be reproduced. When $x\rightarrow \ 0$ $\zeta
(x)=O(x^4)$, so the zero-point contributions to the scalar density of
nucleons and dibaryons, which have the opposite signs, are comparable for $%
4g_\sigma ^4/m_N\approx h_\sigma ^4/m_D$. The dibaryon effects become large
for $h_\sigma /(2g_\sigma )\approx 0.5(4m_D/m_N)^{1/4}\approx 0.84$. The
greater dibaryon mass, the greater the upper limit to the ratio $h_\sigma
/(2g_\sigma )$. This effect is seen in Fig.1.

The saturation curve is shown in Fig.2 and the effective nucleon mass
dependence on the Fermi wavenumber $k_F$ is shown in Fig.3 for $h_\sigma
/(2g_\sigma )=0.6$ in case of the H-particle. EOS in RHA is softer than in
MFT. The contributions of the vacuum zero-point fluctuations of nucleons and
dibaryons partially cancel each other, so the inclusion of the dibaryons
makes the EOS stiffer. In Fig.2, we see that the dashed curves corresponding
to RHA with dibaryons lie above the solid lined corresponding to the RHA
with no dibaryons. The same effect is seen on Fig.3. In MFT, the nucleon
effective mass decreases with the density faster then in RHA. Due to the
partial compensation of the nucleon and dibaryon contributions to the vacuum
scalar density, the dashed lines lie below the solid lines.

In Fig.4 we show the critical densities for occurrence of dibaryons H(2220),
d'(2060), and d$_1$(1920) in MFT and RHA in nuclear and neutron matter
versus the ratio $h_\sigma /(2g_\sigma )$ for $h_\omega /h_\omega ^{max}=1$, 
$0.8$, and $0.6$ where $h_\omega ^{max}=h_\sigma m_\omega /m_\sigma $ is the
maximum value for the $\omega $-meson coupling constant with dibaryons at
which the inequality (11) is violated ($h_\omega /h_\omega ^{max}=1$
corresponds to $a_s^2=0$ and $h_\omega /h_\omega ^{max}=0.8$ and $0.6$
correspond to $a_s^2<0$). In RHA, dibaryons occur at higher densities. The
coupling constant $h_\omega $ determines the energy of dibaryons in the
positive $\omega $-meson mean field. The greater the $h_\omega $, the
greater the density is required to make production of dibaryons
energetically favorable. This effect is seen in Fig.4: The solid lines $%
h_\omega /h_\omega ^{max}=1$ lie above the long-dashed and dashed lines $%
h_\omega /h_\omega ^{max}=0.8$ and $0.6$, respectively.

When the dibaryon matter is unstable against compression, production of
dibaryons with increasing the density results to instability of neutron
stars with subsequent phase transition into the quark matter and conversion
of neutron stars into quark stars, strange stars, or black holes. In such a
case, the maximum masses of neutron stars are determined by the mass and the
coupling constants of the mesons with the lightest dibaryon. In Fig.5 we
show the minimal neutron star masses in which dibaryons can occur.

The MFT and RHA EOS for neutron matter at supranuclear densities are matched
smoothly with the BBP EOS \cite{BBP} at densities $\rho _{drip}<\rho
<0.8\rho _0$ where $\rho _{drip}=4.3$ $10^{11}$g/cm$^3$ and then with the
BPS EOS \cite{BPS} at densities $\rho <\rho _{drip}$. The maximum neutron
star masses are sensitive to the value of the equilibrium Fermi wavenumber.
If we chose $k_F=1.42$ fm$^{-1}$ instead of $k_F=1.3$ fm$^{-1}$, the maximum
masses in MFT (with no dibaryons) are reduced from $3M_{\odot }$ down to $%
2.6M_{\odot }$ \cite{Wal}. The choice $k_F=1.3$ fm$^{-1}$ provides less
stringent (therefore more conservative) constraints to the meson-dibaryon
coupling constants. We do not show the results for the d$_1$(1920) dibaryon,
since its condensation starts at a density $\rho \approx $\ $\rho _0$
providing conversion to quark stars of neutron stars with very low masses $%
M<0.2M_{\odot }$.

In Fig. 6 we show the parameter space for the coupling constants of
dibaryons with the mesons. Our discussion and the validity of our arguments
are restricted to the region $a_s^2<0$ in which the dibaryon matter is
unstable against compression. The requirement of stability of the normal
nuclear matter at the saturation density allows to get constraints to the
coupling constants. The corresponding curves (straight lines in the MFT
case) marked by arrows with the white ends restrict from below the parameter
space of the coupling constants. The dotted line in the MFT case, which
refers to the d$_1$(1920) dibaryon, is very close to the dashed-dotted line $%
a_s^2=0$. For low values $h_\sigma $ the dotted line lies above the line $%
a_s^2=0$. In such a case, the dibaryon matter unstable against compression
cannot exist. The window in the parameter space for the unstable dibaryon
matter for d$_1$(1920) is, however, much greater for RHA.

If a conservative assumption is used, namely, that pulsars with a mass $%
1.3M_{\odot }$ are ordinary neutron stars, the constraints to the meson
coupling constants with dibaryons can further be improved. The corresponding
curves (straight lines in the MFT case) are shown on Fig.6. We see that the
dotted and dashed lines lie above or very close to the line $a_s^2=0$. It
means that for d$_1$(1920) and d'(2060) dibaryons, the dibaryon matter
unstable against compression cannot exist (owing to a very small window for
d'(2060) at higher values of the $h_\sigma $). For the H-particle, there is
a window in the parameter space between the line $a_s^2=0$ and the solid
curves marked by arrows with the black ends, which corresponds to the
dibaryon matter unstable against compression.

The H-particle interaction were studied in the non-relativistic quark
cluster model \cite{Str,Oka} successful in describing the $NN$-phase shifts.
The coupling constants of the mesons with the H-particle can be fixed by
fitting the depth and the position of the minimum of the HH-adiabatic
potential \cite{Sak} to give $h_\omega /(2g_\omega )=0.89$ and $h_\sigma
/(2g_\sigma )=0.80$. These values are marked on Fig.6 with a cross. These
estimates are used in the MFT-calculations \cite{Fae,Faes}. They correspond
to the unstable dibaryon matter and are in the allowed region of the
parameter space for the H-particle. The energetically favorable compression
of the H-matter can lead to the formation of the absolutely stable strange
mater \cite{Wit}, producing conversion of neutron stars to strange stars 
\cite{Oli}.

MFT and RHA EOS both are very stiff. In the soft models like the Reid one
(for a review of the nuclear matter models see \cite{Sha}), the central
density of neutron stars is much greater than in stiff models. Respectively,
conditions for occurrence of new forms of nuclear matter are more favorable.
From Figs. 5 and 6, we see that the softer RHA EOS produces lower upper
limits to the neutron star masses and, respectively, more stringent
constraints to the meson-dibaryon coupling constants as compared to the
stiffer MFT EOS, despite in RHA dibaryons occur at higher densities (see
Fig.4). One can assume that this effect is of the general validity and that
softer EOS like the Reid one give more stringent constraints to the
meson-dibaryon coupling constants. We consider therefore the constraints
given in Fig.6 as the conservative ones.

In the conclusion, we showed that the hypothesis of instability of the
dibaryon matter against compression is in contradiction with the hypothesis
that pulsars of a mass $1.3M_{\odot }$ are ordinary neutron stars for
dibaryons d$_1$(1920) and d'(2060). This conclusion is valid for all narrow
dibaryons with the same quantum numbers in the same mass range. The
H-particle is sufficiently heavy, its condensation starts at higher
densities, respectively, constraints to the meson-dibaryon coupling
constants are not much stringent, allowing a possibility for the H-particle
to form a condensate unstable against compression with subsequent formation
of the strange matter. The meson coupling constants with dibaryons d$_1$%
(1920) and d'(2060) should obey the inequality (11). The meson coupling
constants with the H-particle should lie above the solid curves on Fig.6.%
\vspace{0.5cm}

The authors are grateful to B. V. Martemyanov for discussions of the
results. One of the authors (M. I. K.) acknowledges hospitality of Institute
for Theoretical Physics of University of Tuebingen, Alexander von Humboldt
Stiftung for support with a Forschungsstipendium and DFG-RFBR for
Grant No. Fa-67/20-1.

\newpage\

\newpage\ 

\begin{center}
{\bf Figure captions}
\end{center}

\vspace{0.5cm}

{\bf Fig.1} The effective nucleon mass (a), the asymmetry coefficient (b),
the incompressibility of nuclear matter (c) at the saturation density, and
the coupling constants $C_s^2=g_s^2(m_N/m_s)^2$ and $C_\omega ^2=g_\omega
^2(m_N/m_\omega )^2$ in the RHA versus the ratio $h_\sigma /(2g_\sigma )$
between the $\sigma $-meson coupling constants with dibaryons and nucleons.
The MFT results are shown for comparison. The coupling constants are fixed
by fitting the minimum and depth of the energy per baryon number at the
saturation density of nuclear matter. The solid, dashed, and dotted curves
correspond to the dibaryons H(2220), d$^{\prime }$(2060), and d$_1$(1920).
The reasonable description of the properties of the nuclear matter at the
saturation density is possible for $h_\sigma /(2g_\sigma )<0.8$.%
\vspace{0.5cm}

{\bf Fig.2} Saturation curve for nuclear matter in RHA with no dibaryons
(solid line) and in RHA with inclusion of the H-dibaryon (dashed line) for $%
h_\sigma /(2g_\sigma )=0.6$.\vspace{0.5cm}

{\bf Fig.3} The effective nucleon mass versus the fermi momentum of nucleons
in the nuclear matter (upper curves) and in the neutron matter (lower
curves). The solid lines correspond to RHA with no dibaryons, the dashed
lines correspond to RHA with inclusion of the H-dibaryon.\vspace{0.5cm}

{\bf Fig.4} Critical densities for occurrence of the dibaryons H(2220),
d'(2060), and d$_1$(1920) in nuclear and neutron matter in MFT and RHA
versus the $\sigma $-meson coupling constant with dibaryons for $h_\omega
/h_\omega ^{max}=1$, $0.8$, and $0.6$ (the solid, longed-dashed, and dashed
lines, respectively), where $h_\omega ^{max}=h_\sigma m_\omega /m_\sigma $
is the maximum value for the $\omega $-meson coupling constant with
dibaryons at which the dibaryon matter is unstable against compression (see
the text).\vspace{0.5cm}

{\bf Fig.5} The minimum neutron star masses, in which the dibaryon formation
becomes energetically favorable, versus the $\sigma $-meson coupling
constant with dibaryons for $h_\omega /h_\omega ^{max}=1$, $0.8$, and $0.6$
(the solid, longed-dashed, and dashed lines, respectively) where $h_\omega
^{max}$ is the maximum value for the $\omega $-meson coupling constant at
which the dibaryon matter is unstable against compression ($a_s^2<0$). The
results are given in MFT and RHA and for two dibaryons H(2220) and d'(2060).
The minimum neutron star masses for the d$_1$(1920) are very small ($%
<0.2M_{\odot }$).\vspace{0.5cm}

{\bf Fig.6} Parameter space for the coupling constants of the $\sigma $- and 
$\omega $-mesons with dibaryons in MFT and RHA. The dashed-dotted line $%
a_s^2=0$ divides the parameter space into two parts. The upper left part of
the parameter space corresponds to the dibaryon matter stable against
compression (square of the sound velocity is positive $a_s^2>0$), the lower
right part corresponds to the nuclear matter unstable against compression ($%
a_s^2>0$). The solid, dashed, and doted curves constrain from below the
regions in which the dibaryon formation is energetically not favorable in
ordinary nuclei (curves marked by arrows with white ends) and at the density
equal to the central density of a $1.3M_{\odot }$ mass neutron star (curves
marked by arrows with black ends). The cross refers to the H-particle
coupling constants with the mesons, determined from the adiabatic potential 
\cite{Sak}.

\end{document}